\documentstyle[12pt]{article}
\newcommand\be{\begin{equation}}
\newcommand\ee{\end{equation}}
\newcommand\bea{\begin{eqnarray}}
\newcommand\eea{\end{eqnarray}}
\newcommand\ket[1]{|#1\rangle}
\newcommand\bra[1]{\langle #1|}
\newcommand\braket[2]{\langle #1|#2\rangle}
\newcommand{\fatalpha}{{\bf \alpha \kern -0.44em \alpha}}
\newcommand{\fatsigma}{{\bf \sigma \kern -0.54em \sigma}}
\newcommand{\tpchi}{{\bf \chi \kern -0.35em \chi}}
\newcommand{\llambda}{{\bf \lambda \kern -0.45em \lambda}}

\bibliography{plain}
\title{\bf Pseudo-Hermitian continuous-time quantum walks }\vspace{20mm}
\author{ S. Salimi
  \thanks{Corresponding author:  E-mail addresses:
  shsalimi@uok.ac.ir}, \
  A. Sorouri
  \thanks{E-mail: a.sorouri@uok.ac.ir}
 \\ {\small Department of Physics,
University of Kurdistan, P.O.Box 66177-15175 , Sanandaj, Iran.}}
\pagebreak

\vspace{20mm}
\begin{document}
\maketitle \vspace{15mm}
\newpage
\begin{abstract}
In this paper we present a model exhibiting a new type of
continuous-time quantum walk (as a quantum mechanical transport
 process) on networks, which is described by a non-Hermitian Hamiltonian
 possessing a real spectrum. We call it pseudo-Hermitian continuous-time quantum walk.
 We introduce a method to obtain the probability distribution of walk on any vertex
 and then study a specific system. We observe that the probability
 distribution on certain vertices increases compared to that of the
 Hermitian case. This formalism makes the transport process faster and
 can be useful for search algorithms.


{\bf PACs Index: 03.65.Ud }
\end{abstract}

\vspace{70mm}
\newpage
\section{Introduction}
In recent years, the studies of quantum random walks have suggested
that they may display different behavior from their classical
counterpart \cite{adz, fg}. One of the promising features of quantum
random walks is that they provide an intuitive framework on which
one can build novel quantum algorithms. Since many classical
algorithms can be formulated in terms of random walks, one can hope
that some of them may be translated into quantum algorithms which
run faster than their classical counterparts. However, Shenvi et.
al.\cite{Shenvi} have shown that the quantum search algorithm can be
derived from a certain kind of quantum random walks. Also, Childs
et. al. \cite{child} have presented a general approach to the Grover
problem using a continuous-time quantum walk (CTQW) on a graph.

Quantum walk is generally divided into two standard versions:
discrete-time quantum walk (DTQW) and CTQW \cite{kempe,Hoyer,
Kempf,Chandrashekar, Belton, Sanders, Strauch, Santha}. In the past
years,  CTQW has been studied on the line \cite{jas1,jas3, konno},
distance regular graphs~\cite{jas2}, $n$-cube \cite{moore}, star
graphs \cite{ssa1, xp1}, quotient graphs~\cite{SS},  circulant
 Bunkbeds~\cite{LRS}, decision trees~\cite{FG, konno2}, odd graphs
 \cite{shs}, Apollonian network~\cite{XL}, and dendrimers \cite{OMV}.
 In all of the above-mentioned works,  CTQW  has been  described by  Hermitian Hamiltonians on networks.
 In this paper we describe a model exhibiting a new type of CTQW (as a quantum mechanical transport
 process) on networks, which is described by a non-Hermitian Hamiltonian
 possessing a real spectrum. In recent years, we have witnessed a growing interest in the
 study of $\mathcal{PT}$-symmetric Hamiltonian. These operators are
 specific by  their  $\mathcal{PT}$-symmetry, i.e. $[{\mathcal{PT}},
 H]=0$, where $\mathcal{P}$ is space reflection and  $\mathcal{T}$
is time reversal, and can be covered by a broad class of
pseudo-Hermitian operators which satisfy the following operator
equation
$$
H^\dagger= \Theta H\Theta^{-1},
$$
where the operator $\Theta$ is required to be Hermitian, invertible
and bounded \cite{cmb1}-\cite{a8}. {\bf{One of the first application
of the apparently non-Hermitian Hamiltonian with real spectra
appeared in nuclear physics \cite{scholtz} }. Also, in the recent
works, attention has been paid to the extension of the
phenomenological scope as well as the practical feasibility of the
use of the Hamiltonians $H$ which are only made Hermitian after the
introduction of sophisticated, Hamiltonian-dependent \textit{ad hoc}
metrics \cite{Znojil1, Znojil2}, and  studied for non-Hermitian
matrices Hamiltonian  $2\times 2$ and  $3\times 3$ on Hilbert spaces
with dimensions $2$ and $3$ \cite{Znojil3, Znojil4}, respectively. }

 Here, we consider that kind of
the graphs whose  Hamiltonians are pseudo-Hermitian. Using
pseudo-Hermitian quantum mechanics, we obtain the probability
distribution (transition probability) of walk on any vertex. The
results show that the probability distribution on certain vertices
increases in comparison with that of  Hermitian CTQW. Therefore, we
can increase the probability distribution on any arbitrary vertex.
Noting the fact that the Grover search algorithm is a quantum random
walk search algorithm \cite{Shenvi, miklos, potocek, gabris,Avatar},
pseudo-Hermitian quantum walk is a better candidate than Hermitian
quantum walk for search algorithm. Also, by attention to increasing
and decreasing of probability distribution on any arbitrary vertex,
this formalism gives faster transport on networks than that of the
Hermitian case. This paper is organized as follows: In Sec. $2$ we
first review CTQW on general graphs. Then, we describe the
pseudo-Hermitian formalism and define pseudo-Hermitian CTQW (PHCTQW)
and obtain probability distribution for PHCTQW on general graphs. In
Sec $3$ we derive  PHCTQW for an example. Conclusion is given in the
last part, Sec $4$.

\section{Pseudo-Hermitian continuous-time quantum walk}
CTQW was introduced by Farhi and Gutmann \cite{fg} as a quantum
mechanical transport process on discrete structures, called
generally graphs. A graph is a pair of two sets, $G=(V,E)$, where
$V$ is a non-empty set and $E$ is a subset of $\{(i, j): i, j\in V ,
i\neq j \}$. The elements of $V$ and $E$ are called vertices and
edges, respectively. The adjacency matrix of $G$ is defined as
\begin{equation}
A_{i j}\;=\;\cases{1 & if $\;(i, j)\in V $,\cr 0 &
otherwise\cr}\qquad \qquad (i, j \in V).
\end{equation}
We assume that  states $\ket{k}$, associated with localized
excitation at vertex $k$, form an orthonormal basis set and span the
whole accessible Hilbert space. The Hamiltonian of  quantum walk is
defined as the Laplacian of the graph, $H=A-D$, where $D$ is a
diagonal matrix with entries $D_{jj}=deg(j)$, where $deg(j)$ is the
degree of the vertex $j$ which is the number of proper edges
incident on $j$. Time evolution of quantum walk on graphs obtained
by using the Schr\"{o}dinger equation:
\begin{equation}\label{sh1}
i\hbar\frac{d}{dt}\ket{\psi(t)} = H\ket{\psi(t)}.
\end{equation}
The solution of this equation is $\ket{\psi(t)} = e^{-iHt}
\ket{\psi(0)}$, where we assume $\hbar = 1$,  and $\ket{\psi(0)}$ is
the initial  wave function amplitude of the particle. The transition
probability of quantum walk at vertex $k$ at time $t$ is given by
\begin{equation}
p^q_k(t)=|\langle k|\psi(t)\rangle|^2,
\end{equation}
where $"q"$ stands for quantum walk. Also, by using Kolmogorov's
equation one can obtain transition probability for classical random
walk  as
\begin{equation}\label{ko1}
 p^c_{k}(t) = \langle k|e^{-Ht} \ket{\psi(0)},
\end{equation}
where $"c"$ refers to classical walk. To solve  equations
(\ref{sh1}) and (\ref{ko1}) exactly, we need to know  all the
eigenvalues and eigenvectors of the Hermitian Hamiltonian $H$. Let
$E_n$ and $\ket{q_n}$ be the $n$-th eigenvalue and the corresponding
orthonormalized eigenvector of $H$, respectively. Hence, the
classical and quantum mechanical transition probabilities
 at vertex $k$ at  time $t$ are given by
\begin{equation}
p^q_k(t)=\sum_{n,l}e^{-it(E_n-E_l)}\langle k|q_n\rangle\langle
q_n|\psi_{0}\rangle\langle k|q_l\rangle\langle q_l|\psi (0)\rangle,
\end{equation}
\begin{equation}
p^c_k(t)=\sum_{n}e^{-tE_n}\langle k|q_n\rangle\langle q_n|\psi
(0)\rangle,
\end{equation}
respectively.

Here, we consider a model exhibiting a new type of quantum
mechanical transport processes on networks described by a
non-Hermitian Hamiltonian possessing a real spectrum.

Let $\mathcal{H}$ be a Hilbert space and $H :
\mathcal{H}\longrightarrow \mathcal{H}$  be a diagonalizable linear
Hamiltonian operator. $H$ is said to be pseudo-Hermitian if there
exists a positive-definite operator $\Theta :
\mathcal{H}\longrightarrow \mathcal{H}$  such that \cite{a8}
\begin{equation}\label{her1}
H^\dagger=\Theta H \Theta^{-1}.
\end{equation}
It can be shown that $H$ is a Hermitian operator with respect to
some positive-definite inner product $\langle.,.\rangle_+$ on
$\mathcal{H}$ (which is generally different from its defining inner
product $\langle.|.\rangle$ ). A specific choice for
$\langle.,.\rangle_+$ is $\langle.|\Theta .\rangle$. Therefore, one
can consider the set of the $H$  eigenvectors  as a basis for
$\mathcal{H}$. If $E_n$  and $\ket{\psi_n}$ are the $n$-th
eigenvalue and the corresponding eigenvector of $H$, respectively,
we have
\begin{equation}\label{eigen}
H\ket{\psi_n}=E_n\ket{\psi_n}.
\end{equation}
One can construct another basis $\{\ket{\phi_n}\}$ of $\mathcal{H}$
satisfying
\begin{equation}\label{compl}
H^\dagger\ket{\phi_n}=E_n\ket{\phi_n}, \quad \langle
\phi_n|\psi_m\rangle=\delta_{nm}, \quad \sum_n |\psi_n\rangle
\langle \phi_n|=1.
\end{equation}
In other words, the set $\{\ket{\psi_n}, \ket{\phi_n}\}$ forms a
biorthonormal system, and we have
\begin{equation}
H=\sum_n E_n |\psi_n\rangle \langle \phi_n|, \qquad H^\dagger=\sum_n
E_n |\phi_n\rangle \langle \psi_n|.
\end{equation}
Therefore, by using Eq.(\ref{her1}) one can define the pseudo-metric
operator $\Theta$   associated with the operator $H$ as \cite {a8}
\begin{equation}\label{metric}
\Theta=\sum_n |\phi_n\rangle\langle\phi_n |, \qquad
\Theta^{-1}=\sum_n |\psi_n\rangle\langle\psi_n |.
\end{equation}
Using the inner product $\langle.,.\rangle_+$, we can introduce a
new vector space, called physical Hilbert space
${\mathcal{H}}_{{phys}}$. This space is actually the same as
$\mathcal{H}$, but with different inner product.

To conserve the total  transition probability of PHCTQW,  we must
consider time evolution of the vector state ($\ket{\psi(t)}$)
determined by the  Schr\"{o}dinger equation in physical Hilbert
space ${\mathcal{H}}_{{phys}}$ with respect to the inner product
$\langle.,.\rangle_+$. In this case, to solve  the Schr\"{o}dinger
equation (\ref{sh1}) for PHCTQW we have
\begin{equation}\label{sh2}
\ket{\psi(t)} = \sum_ne^{-iE_nt}\ket{\psi_n}
\braket{\phi_n}{\psi_0},
\end{equation}
where we use the completeness relations of the basis in Eq.
(\ref{compl}). The norm of the state should be conserved, so
$$
\braket{\psi(t)}{\psi(t)}_+=\sum_{m,n}e^{-i(E_n-E_m)t}\braket{\psi_m}{\Theta\psi_n}\braket{\phi_m}{\psi_0}\braket{\psi_0}{\phi_n}
$$
\begin{equation}\label{prob1}
=
\sum_{n}\braket{\phi_n}{\psi_0}\braket{\psi_0}{\phi_n}=\braket{\psi_0}{(\sum_n
\ket{\phi_n}\bra{\phi_n})|\psi_0}=\braket{\psi_0}{\Theta\psi_0}=1,
\end{equation}
where
$\ket{\psi_0}=\frac{\ket{\psi(0)}}{(\braket{\psi(0)}{\psi(0)}_+)^{1/2}}$
and we used the definition of $\Theta$ in Eq.(\ref{metric}).
Therefore, by using Eqs.(\ref{sh2}) and (\ref{prob1}) transition
probability of quantum  walk at the vertex $k$ and the time $t$ for
the state $\ket{\psi(t)}$ is given by
\begin{equation}\label{sqprob}
p_{k}^{sq}(t)=|\braket{k}{\Omega|\psi(t)}|^2,
\end{equation}
{\bf{where $\Omega$ need not in general be Hermitian and is defined
as $\Theta=\Omega^{\dagger}\Omega$ \cite{Znojil5,
Znojil6,Znojil7,Znojil8,Znojil9}, but in this paper we consider
$\Omega$ to be a Hermitian operator and defined as
$\Omega=\sqrt{\Theta}$ ; and $"sq"$ stands for Pseudo-Hermitian
quantum walk.}} To obtain the operator $\Omega$ it is necessary to
have the eigenvalues and eigenvectors of operator $\Theta$. Denoting
the eigenvalues and eigenvectors of $\Theta$ by $\epsilon_n$ and
$\ket{\epsilon_n}$, respectively, we have
\begin{equation}
\Theta\ket{\epsilon_n}=\epsilon_n\ket{\epsilon_n}, \quad
\braket{\epsilon_m}{\epsilon_n}=\delta_{mn}, \quad \sum_n
\ket{\epsilon_n}\bra{\epsilon_n}=1.
\end{equation}
These in turn imply
\begin{equation}\label{rho}
\Omega=\sum_n\sqrt{\epsilon_n}\ket{\epsilon_n}\bra{\epsilon_n}.
\end{equation}
\section{Example}

As an example, we consider Pseudo-Hermitian continuous-time quantum
walk on the graph in Fig.1. As can be seen, in this example vertex
$3$ has preference to other vertices. The Hamiltonian $H=D-A$, where
$D$ is the diagonal matrix defined before and $A$ is adjacency
matrix, is non-Hermitian possessing real spectrum which satisfies
Eq.(\ref{her1}). Considering the graph, $H$ is
\begin{equation}
H=\left(
  \begin{array}{ccc}
    2 & -1 & -1 \\
    -1 & 1 & -1 \\
    -1 & 0 & 2 \\
  \end{array}
\right).
\end{equation}

The eigenvalues and eigenvectors (see Eqs.(\ref{eigen}) and
(\ref{compl})) of $H$ and $H^\dagger$ can easily be calculated,
$$
E_1=0, \ E_2=2, \ E_3=3, \quad \ket{\psi_1}=\frac{1}{\sqrt{6}}\left(
                                              \begin{array}{c}
                                                2 \\
                                                3 \\
                                                1 \\
                                              \end{array}
                                            \right), \  \ket{\psi_2}=\frac{1}{\sqrt{2}}\left(
                                              \begin{array}{c}
                                                0 \\
                                                -1 \\
                                                1 \\
                                              \end{array}
                                            \right), \  \ket{\psi_3}=\frac{1}{\sqrt{3}}\left(
                                              \begin{array}{c}
                                                -1 \\
                                                0 \\
                                                1 \\
                                              \end{array}
                                            \right),
$$
\begin{equation}
\ket{\phi_1}=\frac{1}{\sqrt{6}}\left(\begin{array}{c}
                                                1 \\
                                                1 \\
                                                1 \\
                                              \end{array}
                                            \right), \  \ket{\phi_2}=\frac{1}{\sqrt{2}}\left(
                                              \begin{array}{c}
                                                1 \\
                                                -1 \\
                                                1 \\
                                              \end{array}
                                            \right), \  \ket{\phi_3}=\frac{1}{\sqrt{3}}\left(
                                              \begin{array}{c}
                                                -2 \\
                                                1 \\
                                                1 \\
                                              \end{array}
                                            \right).
\end{equation}

Using Eq.s(\ref{metric}) and (\ref{rho}) the operators $\Theta$ and
$\Omega$ are obtained as
\begin{equation}\label{rho1}
\Theta=\left[ \begin {array}{ccc}
2&-1&0\\\noalign{\medskip}-1&1&0\\\noalign{\medskip}0&0&1\end
{array} \right], \quad \Omega=\left[ \begin {array}{ccc}
\frac{3}{5}\sqrt {5}&-\frac{1}{5}\sqrt
{5}&0\\\noalign{\medskip}-\frac{1}{5}\sqrt{5}&\frac{2}{5}\sqrt
{5}&0\\\noalign{\medskip}0&0&1\end {array} \right].
\end{equation}
Therefore, the solution of the Schr\"{o}dinger  equation (\ref{sh2})
is
\begin{equation}
\ket{\psi(t)}=\frac{1}{\sqrt{12}}\ket{\psi_1}+\frac{1}{\sqrt{2}}e^{-i2t}\ket{\psi_2}-\frac{2}{\sqrt{6}}e^{-i3t}\ket{\psi_3},
\end{equation}
where we consider $\ket{\psi(0)}=\ket{1}$ as the starting point of
the walk. Then using Eqs.(\ref{sqprob}) and (\ref{rho1}) we obtain
the transition probability as
\begin{equation}
p_{k}^{sq}(t) = \left\{\begin{array}{ccc}
\frac{1}{40} [18 + 8 \cos(t) + 2 \cos(2 t)+ 8 \cos(3 t) ] & \mbox{ for $k=1$ } \\
\frac{1}{90} [17+12\cos(t) - 12\cos(2 t)-8\cos(3t)] & \mbox{ for $k=2$ } \\
\frac{1}{72} [26-24 \cos(t) + 6 \cos(2 t)- 8 \cos(3 t) ]  &
\mbox{for $k=3$}.
            \end{array}\right.
\end{equation}
Fig.2 shows the transition  probability on the network in Fig.1,
where the starting point of the walk is node $\ket{1}$. Fig.2(a)
illustrates the probability of returning to the starting point ($
\ket{1}$). As can be seen, in returning to the starting point, the
walk spends most of the time with the probability between 0.3 and
0.5. However, in the Hermitian version (CTQW on complete graph $K_3$
) the probability on vertex 1 is homogeneous for all time. Fig.2(b)
shows that the probability of observing the walk in vertex $2$ is
more than that of the Hermitian version. The same happens for vertex
3, but ,as shown in Fig.2(c), the difference between the
probabilities for the two cases (Hermitian and non-Hermitian) at
vertex 3 is much more than that at vertex 2. In fact, one can apply
the PHCTQW for increasing the probability of observing the walk at
an arbitrary vertex, which is very useful to make the transport
faster on a network.

Now, let vertex $2$ be the starting point of the walk. The
transition probability is obtained as
\begin{equation}
p_{k}^{sq}(t) = \left\{\begin{array}{ccc}
\frac{1}{10} [3 + 2 \cos(t) - \cos(2 t)-2 \cos(3 t) ] & \mbox{ for $k=1$ } \\
\frac{1}{45} [14+6\cos(t) + 12\cos(2 t)+4\cos(3t)] & \mbox{ for $k=2$ } \\
\frac{1}{36} [14-12\cos(t) -6 \cos(2 t)+4 \cos(3 t) ]  & \mbox{for
$k=3$}.
\end{array}\right.
\end{equation}
Fig.3 shows the behavior of $p_{k}^{sq}(t)$ with respect to time
$t$. Fig.3(a) shows the probability distribution of finding the walk
at vertex 1. Although the distribution is not as homogeneous as in
the Hermitian case, the value of the probability has not generally
changed so much. The return probability distribution to vertex 2
(starting point) is illustrated in Fig.3(b). As can be seen, it
changes periodically due to the special property of the graph. In
Fig.3(c) we see that the transition probability of the walk is
increased in comparison with its transition probabilities on other
vertices.

In this example we saw that the non-Hermitian property of the
Hamiltonian is responsible for increasing the probability of
observing a particle on any vertex.

{\bf{Also, one can generalize this method to obtain PHCTQW on the
graph shown in Fig.4. Fig.4 illustrates a cycle graph with $n$
vertices such that the edge between vertex $n-1$ and $n$th is
directed, in other words, vertex $n$ has preference to the other
vertices. As an example, we consider the above-mentioned graph with
$4$ vertices. The Hamiltonian is given by
\begin{equation}
H=\left(
    \begin{array}{cccc}
      2 & -1 & 0 & -1 \\
      -1 & 2 & -1 & 0 \\
      0 & -1 & 1 & -1 \\
      -1 & 0 & 0 & 2 \\
    \end{array}
  \right).
\end{equation}
The eigenvalues and eigenvectors (see Eqs.(\ref{eigen}) and
(\ref{compl})) of $H$ and $H^\dagger$ can be obtained as,
$$
E_1=0, \ E_2=2, \ E_3=\frac{1}{2}(5+\sqrt{5}),\
E_4=\frac{1}{2}(5-\sqrt{5}), \quad
\ket{\psi_1}=\frac{1}{\sqrt{10}}\left(
                                              \begin{array}{c}
                                                2 \\
                                                3 \\
                                                4\\
                                                1 \\
                                              \end{array}
                                            \right), \  \ket{\psi_2}=\frac{1}{\sqrt{2}}\left(
                                              \begin{array}{c}
                                                0 \\
                                                -1 \\
                                                0\\
                                                1 \\
                                              \end{array}
                                            \right),
$$$$
                                            \ket{\psi_3}=\frac{i}{\sqrt{\frac{5}{2}(3+\sqrt{5})}}\left(
                                              \begin{array}{c}
                                                \frac{1}{2}(1+\sqrt{5}) \\
                                                -\frac{1}{2}(1+\sqrt{5}) \\
                                                1 \\
                                                -1\\
                                              \end{array}
                                            \right),\ \ket{\psi_4}=\frac{i}{\sqrt{\frac{5}{2}(3-\sqrt{5})}}\left(
                                              \begin{array}{c}
                                                \frac{1}{2}(1-\sqrt{5}) \\
                                                \frac{1}{2}(-1+\sqrt{5}) \\
                                                1 \\
                                                -1\\
                                              \end{array}
                                            \right),
$$
$$
\ket{\phi_1}=\frac{1}{\sqrt{10}}\left(\begin{array}{c}
                                                1 \\
                                                1 \\
                                                1\\
                                                1 \\
                                              \end{array}
                                            \right), \  \ket{\phi_2}=\frac{1}{\sqrt{2}}\left(
                                              \begin{array}{c}
                                                -1 \\
                                                -1 \\
                                                1\\
                                                1 \\
                                              \end{array}
                                            \right), \  \ket{\phi_3}=\frac{i}{\sqrt{\frac{5}{2}(3+\sqrt{5})}}\left(
                                              \begin{array}{c}
                                                -(1+\sqrt{5}) \\
                                                \frac{1}{2}(3+\sqrt{5}) \\
                                                -1 \\
                                                \frac{1}{2}(3+\sqrt{5})\\
                                              \end{array}
                                            \right),
$$
\begin{equation}
                                             \ket{\phi_4}=\frac{i}{\sqrt{\frac{5}{2}(3-\sqrt{5})}}\left(
                                              \begin{array}{c}
                                                (-1+\sqrt{5}) \\
                                                \frac{1}{2}(3-\sqrt{5}) \\
                                                -1 \\
                                                \frac{1}{2}(3-\sqrt{5})\\
                                              \end{array}
                                            \right).
\end{equation}
In this case, by using Eq.(\ref{metric}), we obtain the operator
$\Theta$ as follows
\begin{equation}
\Theta=\left(
        \begin{array}{cccc}
          \frac{11}{5} & \frac{1}{5} & -\frac{4}{5} & -\frac{4}{5} \\
          \frac{1}{5} & \frac{6}{5} & -\frac{4}{5} & \frac{1}{5} \\
          -\frac{4}{5} & -\frac{4}{5} & \frac{6}{5} & \frac{1}{5} \\
          -\frac{4}{5} & \frac{1}{5} & \frac{1}{5} & \frac{6}{5} \\
        \end{array}
      \right).
\end{equation}
Therefore, one can investigate PHCTQW for this graph as was done for
the previous example. }}

\section{Conclusions}
In this paper we have introduced a model exhibiting a new type of
CTQW on networks, which is described by a non-Hermitian Hamiltonian
possessing a real spectrum. The model is called PHCTQW. Then we
introduced a method to obtain the probability distribution of walk
on any vertex.
 We have applied this model to a network and saw that the probability
 distribution on certain vertices increased compared to its
 Hermitian counterpart. Indeed by choosing a non-Hermitian
 Hamiltonian for CTQW one can increase the probability distribution
 on an arbitrary vertex which can help to make the transport faster and
 can be useful for search algorithms. One can generalize
 this method to perfect quantum state transfer over some special
 networks, which is under investigation now.

{\center{\bf{Acknowledgments}}}

We thank Dr. Najarbashi and Prof. Arkat for useful discussions.

\newpage
{\bf Figure Captions}

{\bf Figure-1:} The edge $(2,3)$ is directed and other edges are
undirected.

{\bf Figure-2:} The probability distribution of the walk on vertices
when starting point is on the vertex 1: (a) return probability of
the walk to the starting point (vertex 1), (b) the probability
distribution of walk on vertex $2$ and (c) the probability
distribution of walk on vertex $3$ for PHCTQW (blue line) and its
Hermitian version (red line).

{\bf Figure-3:} The probability distribution of the walk on vertices
of graph when the starting point of the walk is on the vertex $2$ of
Fig.1: (a) the  probability distribution of walk on vertex $1$ in
terms $t$ , (b) the probability of returning to the starting point
(vertex $2$) and (c) the probability distribution of the walk on
vertex $3$ graph for PHCTQW (blue line ) and  its Hermitian version
(red line).

{\bf{Figure-4: Fig.4 shows a cycle graph such that the vertex $n$
has preference to the other vertices.}}

\end{document}